\documentstyle[12pt]{article}
\def\abstracts#1#2#3{{
        \centering{\begin{minipage}{4.62in}\baselineskip=13pt

        \small
        \centerline{\bf Abstract}
        \vspace*{0.2cm}             
        \parindent=0pt #1\par
        \parindent=18pt #2\par
        \parindent=15pt #3
        \end{minipage} }\par}}
\begin{document}
\vspace*{-2cm}
\hfill \parbox{5.5cm}{ Mainz preprint KOMA-98-41\\
%                     printed: \today }\\
                       }\\
\vspace*{1.5cm}
\centerline{\LARGE \bf High-Temperature Series Analysis
           }\\[0.3cm]
\centerline{\LARGE \bf of the Free Energy and Susceptibility
           }\\[0.3cm]
\centerline{\LARGE \bf of the 2D Random-Bond Ising Model
           }\\[0.8cm]
%\addtocounter{footnote}{-1}
%\renewcommand{\thefootnote}{\arabic{footnote}}
\vspace*{0.2cm}
\centerline{\large {\em Alexandra Roder\/}$^1$, 
                   {\em Joan Adler\/}$^2$ and
                   {\em Wolfhard Janke\/}$^{1,3}$}\\[0.4cm]
\centerline{\large {\small $^1$ Institut f\"ur Physik,
                                Johannes Gutenberg-Universit\"at Mainz,}}
\centerline{        {\small Staudinger Weg 7, 55099 Mainz, Germany }}\\[0.5cm]
\centerline{\large {\small $^2$ Department of Physics, Technion,}}
\centerline{           {\small Haifa 32000, Israel }}\\[0.5cm]
\centerline{\large {\small $^3$ Institut f\"ur Theoretische Physik,
                                Universit\"at Leipzig,}}
\centerline{    {\small Augustusplatz 10/11, 04109 Leipzig, Germany }}\\[2.5cm]
\abstracts{}{
We derive high-temperature series expansions for the free energy and 
susceptibility of the two-dimensional random-bond Ising model with a
symmetric bimodal distribution of two positive coupling strengths $J_1$
and $J_2$ and study the influence of the quenched, random bond-disorder on 
the critical behavior of the model. By analysing the series expansions over 
a wide range of coupling ratios $J_2/J_1$, covering the crossover from 
weak to strong disorder, we obtain 
for the susceptibility with two different methods compelling evidence 
for a singularity of the form
$\chi \sim t^{-7/4} |\ln t|^{7/8}$, as predicted theoretically by
Shalaev, Shankar, and Ludwig. 
For the specific heat our results are less convincing, but still 
compatible with the theoretically predicted log-log singularity. 
}{}
%\vspace*{1.5cm}
%\noindent PACS numbers:
%
\thispagestyle{empty}
\newpage
\pagenumbering{arabic}
%
%---------------------------------------------------------
           \section{Introduction} \label{sec:intro}
%---------------------------------------------------------
%
One of the most-studied variants of the two-dimensional Ising
model is the case of random bonds.
While realizations of Ising models that include randomness
come much closer to approximating reality, they are very
 much harder to study at any level. Even in two dimensions
exact results for random cases (especially for quenched
randomness, which is the realistic situation in many
experiments)   are few and far-between. In fact, the two-dimensional
Ising case is especially difficult because of the marginality of
the Harris criterion \cite{harris} for this model.
This criterion states that quenched randomness is a relevant
(irrelevant) perturbation when the 
critical exponent $\alpha$ of the specific heat of the pure system is 
positive (negative) and therefore when 
$\alpha=0$ as in the two-dimensional Ising 
model the situation is marginal.

Numerous
theoretical investigations [2-7]
% \cite{DD,shalaev1,shalaev2,shankar,ludwig,ziegler}
as well as numerical Monte Carlo 
simulations [8-15]
% \cite{andreichenko1,wang1,wang2,selke0,selke1,talapov1,wiseman,ssli}
and transfer-matrix studies \cite{reis1,reis2}
have addressed the question of whether the 
critical exponents for the two-dimensional
Ising model with quenched, random bond disorder differ from those of the 
pure model.
While a ``majority'' consensus had probably been achieved in favour of
no change, apart from logarithmic 
corrections [3-6] 
% \cite{shalaev1,shalaev2,shankar,ludwig},
no unambiguous numerical study that confirmed the quantitative
predictions of either of the theoretical approaches had been
 made prior to
our recent study of the susceptibility with high-temperature
 series expansions. In a brief note \cite{raj98} we announced the 
confirmation of the theoretical majority consensus 
value of the exponent of the logarithmic correction.
This quantitative determination of the value of the correction exponent
in excellent agreement with the predicted value, using a completely
different numerical approach that in no way depends on random numbers,
provided
additional and incontrovertible support to the previous consensus.
 In the present paper we present 
the coefficients of the
 susceptibility series that we analysed \cite{raj98}
  together with some remarks on 
their derivation, the details of our analysis, and some additional 
results for the specific-heat series. We note that a recent simulation
of the site-diluted
model \cite{selkepreprint} also confirms the log-log prediction of
% \cite{shalaev1,shalaev2,shankar,ludwig}.
[3-6].

In the next section, we define the model and the quantities that are
studied, and in Sec.~3 the theoretical predictions are briefly recalled. 
The series generation is described in Sec.~4, and in Sec.~5 we
 describe
the analysis techniques used. Sec.~6 then presents the results,
 where details of
our analyses for the susceptibility give compelling evidence for a
singularity of the form predicted by Shalaev, Shankar, and 
Ludwig [3-6].
% \cite{shalaev1,shalaev2,shankar,ludwig}. 
In Sec.~7 we close with
a summary of our conclusions and a few final comments.
%
%---------------------------------------------------------
           \section{Model} \label{sec:model}
%---------------------------------------------------------
%
The Hamiltonian of the random-bond Ising model is given by 
\begin{equation}
{\cal H} = - \sum_{\langle ij \rangle} J_{ij} \sigma_i \sigma_j,
\label{eq:H}
\end{equation}
where the spins $\sigma_i = \pm 1$ are located at the sites of a square
lattice, the symbol $\langle ij \rangle$ denotes nearest-neighbor 
interactions,
and the coupling constants $J_{ij}$ are quenched, random variables. As in
most previous studies we consider a bimodal distribution,
\begin{equation}
P(J_{ij}) = x \delta(J_{ij} - J_1) + (1-x) \delta(J_{ij}-J_2),
\label{eq:P}
\end{equation}
of two ferromagnetic couplings $J_1$, $J_2 > 0$. We furthermore specialize
to a symmetric distribution with $x=1/2$, since in this case the exact 
critical temperature $T_c$ can be computed for any positive value of $J_1$ 
and $J_2$ from the (transcendental) self-duality relation 
($k_B = $ Boltzmann constant) \cite{wu}
\begin{equation}
\left(\exp(2 J_1/k_B T_c) - 1 \right) \left(\exp(2 J_2/k_B T_c) - 1 \right) = 2.
\label{eq:self_dual}
\end{equation}
In both, computer simulations and high-temperature series expansion studies,
this exact information simplifies the analysis of the critical behavior
considerably.

The free energy per site is given by,
\begin{equation}
\beta f = -\lim_{V \rightarrow \infty} \frac{1}{V}
           \left[ \ln \left(\prod_{i=1}^V \sum_{\sigma_i= \pm 1} \right)
           \exp(-\beta {\cal H}) \right]_{\rm av} ,
\label{eq:free}
\end{equation}
where $\beta = 1/k_B T$ is the inverse temperature in natural units and 
the bracket $[ \dots ]_{\rm av}$ denotes the average over the quenched,
random disorder, $[ \dots ]_{\rm av} = 
\left( \prod_{\langle ij \rangle} \int d J_{ij}\right)$ $(\dots) P(J_{ij})$.
The internal energy and specific heat per site follow by differentiation
with respect to $\beta$, 
\begin{equation}
e = \partial \beta f/\partial \beta, \qquad
C/k_B = - \beta^2 \partial^2 \beta f/\partial \beta^2.
\label{eq:C}
\end{equation}
In this paper we shall mainly focus on the magnetic susceptibility per 
site $\chi$ which in the high-temperature phase and zero external field
is defined as the $V \longrightarrow \infty$ limit of
\begin{equation}
% NO beta in front!
\chi_V = \left[ \left\langle \left( \sum_{i=1}^V \sigma_i \right)^2 
                    \right\rangle_T/V \right]_{\rm av},
\label{eq:chi}
\end{equation}
where $\langle \dots \rangle_T$ denotes the usual thermal average 
with respect to $\exp(-\beta {\cal H})$.
%
%---------------------------------------------------------
           \section{Theoretical predictions} \label{sec:theory}
%---------------------------------------------------------
%
Let us briefly recall two contradicting theoretical predictions for
the critical behavior of the model (\ref{eq:H}), (\ref{eq:P}). The 
first is based on renormalization-group techniques developed 
by Dotsenko and Dotsenko (DD) \cite{DD}. For the specific heat DD find
close to the transition point a double logarithmic behavior,
\begin{equation}
C(t)\propto \ln(\ln(1/|t|)),
%C(t) = \frac{1}{g_0} \ln \left[1 + \frac{4 g_0}{\pi} 
%\ln \left( \frac{1}{|t|}\right)\right],
\label{eq:C_DD}
\end{equation}
where $t = (T-T_c)/T_c$ denotes the reduced temperature.
For the susceptibility
they obtain in the high-temperature phase ($t \ge 0$)
\begin{equation}
\chi \propto t^{-2} \exp\left[ -a \left(\ln \ln \left(\frac{1}{t}\right)
\right)^2 \right].
\label{eq:chi_DD}
\end{equation}
The second approach by Shalaev, Shankar, and Ludwig (SSL) 
% \cite{shalaev1,shalaev2,shankar,ludwig} 
[3-6] makes use of
bosonisation techniques and the method of conformal invariance. While
the prediction (\ref{eq:C_DD}) for the specific heat 
can be reproduced (which, however, is not undisputed \cite{ziegler}),
SSL derive quite a different behavior for the susceptibility, 
\begin{equation}
\chi \propto t^{-7/4} | \ln t |^{7/8}.
\label{eq:chi_SSL}
\end{equation}
This is the same leading singularity as in the pure case ($J_1 = J_2$), 
but modified by a multiplicative logarithmic correction.

High-precision Monte Carlo simulations and transfer-matrix studies
% \cite{andreichenko1,wang1,wang2,selke0,selke1,talapov1,wiseman,ssli,reis1} 
[8-16]
favor the latter form,
but due to well-known inherent limitations of this method it has been 
impossible to confirm the value of the exponent of the multiplicative 
logarithmic correction in 
(\ref{eq:chi_SSL}) quantitatively. 
Similar problems have been reported in simulation studies of 
other models exhibiting multiplicative logarithmic corrections such as, e.g.,
the two-dimensional 4-state Potts \cite{potts4} and XY \cite{xylogs} model.
We found it therefore worthwhile to investigate this problem yet again
with an independent method. In the following we report high-temperature 
series expansions for the free energy and susceptibility and enquire whether
series analyses can yield a more stringent test
of the theoretical predictions.
%
%---------------------------------------------------------
           \section{Series generation} \label{sec:series}
%---------------------------------------------------------
%
For the generation of the high-temperature series expansions of
the free energy (\ref{eq:free}), and hence the internal energy and 
specific heat, as well as the infinite-volume limit of the 
susceptibility (\ref{eq:chi})
we made use of a program package developed at Mainz originally for 
the $q$-state Potts spin-glass 
problem [23-27]
% \cite{schreider,schrei_reger,lobe,ljb98,lobe_janke}. 
In this application
the spin-spin interaction is
generalized from $\sigma_i \sigma_j$ to $\delta_{\sigma_i,\sigma_j}$ with
$\sigma_i$ being an integer between 1 and $q$, 
and the coupling constants $J_{ij}$ can take the values
$\pm J$ with equal probability. Since here the coupling constants 
also can take negative values, frustration effects play an important role and
the physical properties of spin glasses \cite{potts_glass} are completely 
different than those of the random-bond system. Technically, however, 
precisely the same 
enumeration scheme for the high-temperature graphs can be employed in both
cases. The only difference is in
the last step where the quenched averages over the $J_{ij}$ are
performed. The details of the employed star-graph expansion technique and our
specific implementation are described 
elsewhere [23-25, 27, 29]
% \cite{schreider,schrei_reger,lobe,lobe_janke,roder}. 
Here we only note that
slight modifications of this program package enabled us to 
generate the high-temperature series expansion for $\beta f$ and $\chi$ up to 
the 11th order in $k = 2 \beta J_1$ for
\begin{itemize}
\item hypercubic lattices of arbitrary dimension $d$,
\item arbitrary number of Potts states $q$,
\item arbitrary probability $x$ in the bimodal distribution, and
\item arbitrary ratios $R = J_2/J_1$, characterizing the 
      strength of the disorder.
\end{itemize}
In this paper we shall concentrate on the two-dimensional ($d=2$)
random-bond Ising model ($q=2$) for a symmetric ($x=1/2$) bimodal 
distribution of two positive coupling strengths $J_1$ and $J_2$.
The series coefficients of the free energy and susceptibility expansions
for various coupling-strength ratios $R = J_2/J_1$ are given in 
Tables~\ref{tab:free.series} and \ref{tab:chi.series}.
Of course, in principle it would be also straightforward to adapt the present
program package to more general probability distributions $P(J_{ij})$.
%
%
%---------------------------------------------------------
           \section{Series analysis techniques} \label{sec:analyse}
%---------------------------------------------------------
%
In the literature many different series analysis techniques have been
discussed which, depending on the type of critical singularity at hand, 
all have their own merits and drawbacks \cite{guttmann1}. In the course 
of this work we have tested quite a few of them \cite{roder}. Here, however,
we will confine ourselves to only those which turned out to be the most useful
for our specific problem at hand.

To simplify the notation we denote a thermodynamic function generically by
$F(z)$ and assume that its Taylor expansion around the origin is known 
up to the $N$-th order,
\begin{equation}
F(z) = \sum_{n=0}^N a_n z^n + \dots.
\label{eq:F}
\end{equation}
If the singularity of $F(z)$ at the critical point $z_c$ is of the 
simple form ($z \le z_c$)
\begin{equation}
F(z) \simeq A (1-z/z_c)^{-\lambda} ,
\end{equation}
with $A$ being a constant, then the ratios of consecutive coefficients 
approach for large $n$ the limiting behavior
\begin{equation}
r_n \equiv \frac{a_n}{a_{n-1}} \simeq \left[ 1 + \frac{\lambda-1}{n}\right]
\frac{1}{z_c}.
\label{eq:r_n}
\end{equation}
 From the offset ($1/z_c$) and slope ($(\lambda-1)/z_c$) of this sequence
as a function of $1/n$
both the critical point $z_c$ and the critical exponent $\lambda$ can
be determined. This is the basis of the so-called ratio method \cite{hunter1}. 
If the critical point $z_c$ is known from other sources (in our case
exactly from self-duality), then one may consider biased extrapolants
for the critical exponent,
\begin{equation}
\lambda_n = nr_n z_c - n + 1 ,
\end{equation}
which simply follow by rearranging eq.~(\ref{eq:r_n}). In the following 
this method will be denoted as ``Biased Ratio I''.

If the singularity of $F(z)$ contains a multiplicative logarithmic
correction (as, e.g., in the SSL prediction for $\chi$),
\begin{equation}
F(z) \simeq A (1 - z/z_c)^{-\lambda} |\ln(1 - z/z_c)|^p ,
\label{eq:F_log}
\end{equation}
then one forms the ratios $r_n$ as before, but considers in addition
the auxiliary function \cite{guttmann2}
\begin{equation}
z^{-p^*} (1-z)^{-\lambda} (\ln[1/(1-z)])^{p^*} = \sum_{n=0}^N b_n z^n + \dots,
\end{equation}
and computes the ratios $r_n^* = b_n/b_{n-1}$. Let us first assume that
the critical exponent $\lambda$ of the leading term is known. Then it can be
shown that the sequence $R_n = r_n/r_n^*$ approaches $1/z_c$ with zero slope
in the limit
$n \longrightarrow \infty$, if and only if $p^* = p$. This determines $p$,
if also $z_c$ is known. If $\lambda$ is not known, then one may vary both
exponents until above relation is satisfied. In the following we refer to
this special ratio method as ``Ln-Ratio''.

Another method \cite{adler5,adlerchang} suitable for a singularity of the 
form (\ref{eq:F_log}) is based on Pad\'e approximants \cite{baker1}. 
Here one generates the series expansion for the auxiliary function
\begin{equation}
G(z) = -(z_c - z) \ln(z_c - z) (\frac{F'(z)}{F(z)} -
\frac{\lambda}{z_c - z}) ,
\label{eq:G}
\end{equation}
which can easily be shown to satisfy
\begin{equation}
\lim_{z \rightarrow z_c} G(z) = p .
\end{equation}
If $z_c$ is known, the value of $G(z)$ at $z=z_c$ can be obtained by 
forming Pad\'e approximants,
\begin{equation}
G(z) \approx [L/M] \equiv \frac{P_L(z)}{Q_M(z)} \equiv
\frac{p_0 + p_1 z + p_2 z^2 + \dots + p_L z^L}
{1 + q_1 z + q_2 z^2 + \dots + q_M z^M},
\label{eq:Pade}
\end{equation}
with $L+M \le N-1$. Note that one order of the initial series is lost
due to the differentiation in (\ref{eq:G}).

With a small modification this method can also be applied to a purely
logarithmic singularity of the form
\begin{equation}
F(z) \simeq A |\ln(1 - z/z_c)|^p .
\end{equation}
In this case one defines the auxiliary function
\begin{equation}
G(z) = -(z_c - z) \ln(z_c - z) \frac{F'(z)}{F(z)},
\end{equation}
which again satisfies
\begin{equation}
\lim_{z \rightarrow z_c} G(z) = p.
\end{equation}
The two analysis methods based on Pad\'e approximants will be called
``Ln-Pad\'e''.
%
%---------------------------------------------------------
           \section{Results} \label{sec:results}
\begin{figure}[bhp]
\vskip 7.0truecm
\includegraphics{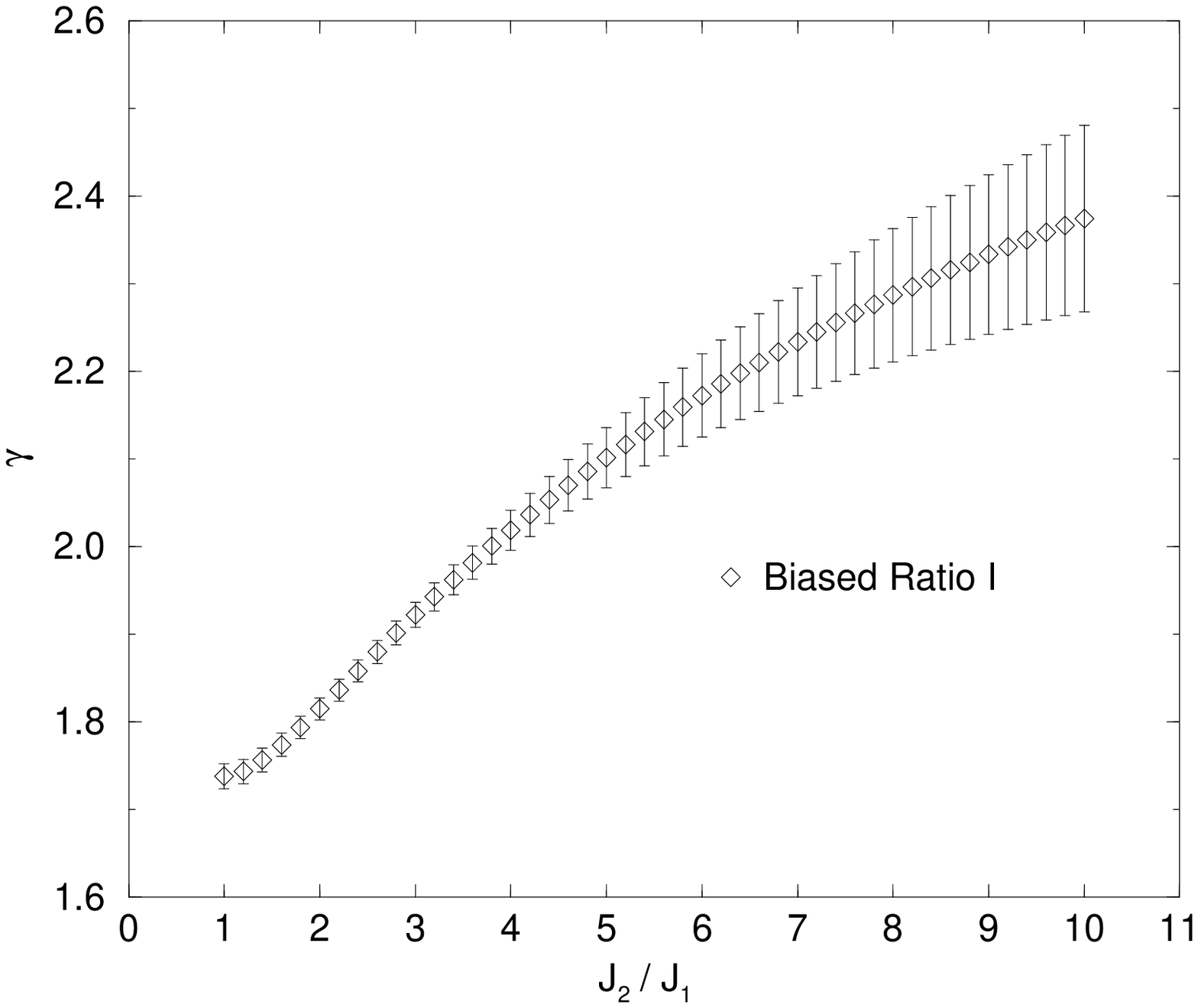}
\caption[a]{{\em Analysis of the susceptibility series assuming a 
singularity of the form $\chi \propto t^{-\gamma}$, using the method 
``Biased Ratio I''. 
}}
\label{fig:chi_DD}
%\end{figure}
%
%\clearpage\newpage
%
%-------------------------------------------------------------------
%                         Figure 2 and 3
%-------------------------------------------------------------------
%
%\begin{figure}[bhp]
\vskip 8.5truecm
\includegraphics{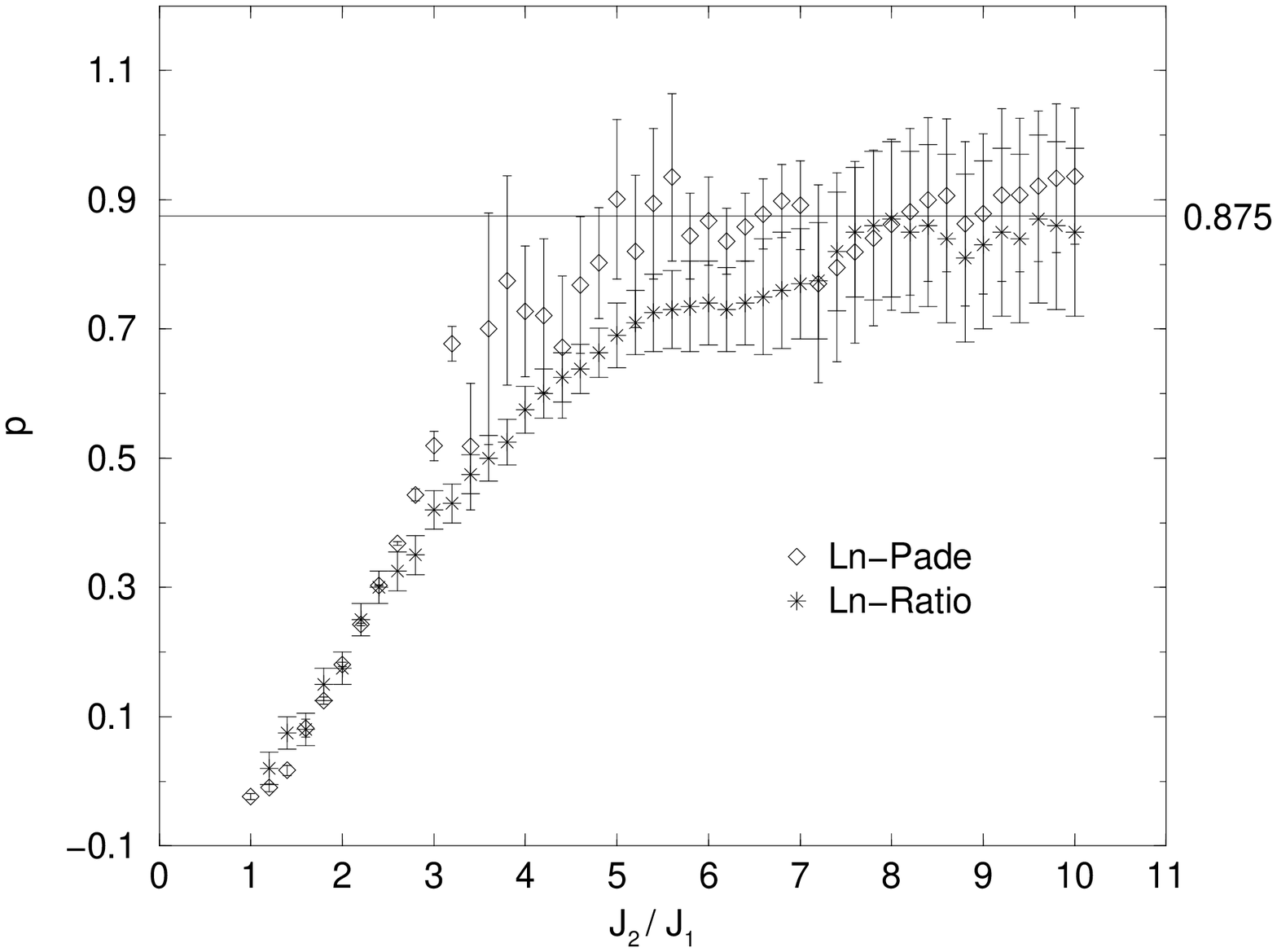}
\caption[a]{{\em Analysis of the susceptibility series assuming a 
singularity of the form $\chi \propto t^{-7/4} |\ln t|^p$, using 
Pad\'e approximants and the ratio method (see text). The horizontal
line at $p = 7/8 = 0.875$ is the theoretical prediction of SSL.
}}
\label{fig:chi_SSL}
\end{figure}

%---------------------------------------------------------
%
\paragraph{Susceptibility:}
In a first step we investigated whether our series expansions for
the susceptibility are consistent with a pure power-law behavior
according to the DD prediction (\ref{eq:chi_DD}) (ignoring the exponentially
small multiplicative correction term). Assuming thus the behavior
$\chi \propto t^{-\gamma}$ and using the method ``Biased Ratio I'' we
obtained the critical exponents $\gamma$ shown in Fig.~\ref{fig:chi_DD}
as a function of $J_2/J_1$.
Here and in the following the error bars are estimated by varying the
length of the series and/or the type of Pad\'e approximants used.
Starting with $\gamma = 1.738 \pm 0.014$ for the pure case ($J_2/J_1=1$),
being consistent with the exact value of $\gamma = 7/4$, we observe
a steady increase to $\gamma = 2.37 \pm 0.11$ for the strongest
investigated disorder ($J_2/J_1=10$). We will argue below that the
apparent crossover from weak to strong disorder is due to the finite
length of our series expansion which naturally has a much more dramatic
influence for weak disorder. At any rate, for strong disorder the
DD prediction of $\gamma=2$ is clearly outside the error margins of the
series analysis estimates.

So far no multiplicative logarithmic corrections were taken into account. 
If the SSL prediction (\ref{eq:chi_SSL}) was correct we would, therefore, 
expect to observe ``effective'' critical exponents which according to
\begin{equation}
\chi \propto t^{-7/4} | \ln t |^{7/8} 
     = t^{-(7/4) [1 + \frac{1}{2} \frac{\ln(|\ln t|)}{\ln(1/t)}]}
\label{eq:chi_eff}
\end{equation}
should indeed be larger than $7/4$. The results in Fig.~\ref{fig:chi_DD}
could thus be well consistent with a critical exponent of $\gamma = 7/4$ in
the presence of a multiplicative logarithmic correction.
 
This possibility suggested a more careful analysis based on the
qualitative form of the SSL prediction (\ref{eq:chi_SSL}). Our series are 
too short to employ a general ansatz with both exponents as free 
parameters. We rather fixed the exponent $\gamma = 7/4$ of the leading term 
to the (predicted) pure Ising model value and enquired if our series
expansions are compatible with the ansatz
\begin{equation}
\chi \propto t^{-7/4} |\ln t |^p,
\end{equation}
and $p=7/8$. Employing the two special methods for this type of
singularity described in Sec.~\ref{sec:analyse} we obtained well
converging results. The resulting estimates for the exponent $p$
are shown in Fig.~\ref{fig:chi_SSL}. We see that the two methods
yield consistent results which start in the pure case ($J_2/J_1=1$)
around $p=0$, as they should do. With increasing disorder the estimates
exhibit again an apparent crossover, until around $J_2/J_1 = 5 - 8$ they
settle at a plateau value in very good agreement with the theoretical
prediction of $p=7/8$. This is the main result of our series analysis.
We claim that compared with previous methods this is thus far the clearest
quantitative confirmation of the SSL prediction (\ref{eq:chi_SSL}). 

As before we attribute the apparent crossover for
intermediate strength of the disorder to the shortness of our series
expansions, i.e., we interpret the crossover as an unavoidable artifact 
of high-temperature series expansion analyses and not as an indication
that the exponent $p$ really is a function of the disorder strength.  
We thus take the view that already a small amount of disorder drives the
system into a new universality class different from the pure case which,
however, only becomes visible in the very vicinity of the transition
point $T_c$ (or $t=0$). This in turn translates into the need of
extremely long series expansions in order to be detectable.

To justify this claim we have investigated a model function simulating
the ``true'' susceptibility ($g_0 \ge 0$), where $g_0$ is a constant that 
depends on the strength of the disorder,
\begin{equation}
\chi_{\rm model} = \dot{t}^{-7/4} 
                   \left[ 1 + \frac{4 g_0}{\pi} \ln (1/\dot{t}) \right]^{7/8},
\label{eq:chi_model}
\end{equation}
with $\dot{t} = (T - T_c)/T$,
which for any $g_0 \ne 0$ reproduces the SSL form (\ref{eq:chi_SSL})
in the limit $T \longrightarrow T_c$ ($\dot{t} = t - t^2 + t^3 + \dots 
\longrightarrow 0$).
Notice the discontinuity in the asymptotic behavior at $g_0=0$.
For any $g_0 \ne 0$ the asymptotic region is 
reached when $\ln(1/t)$ is much larger than $\pi/4 g_0$, i.e., 
for $t \ll \exp(-\pi/4 g_0)$. Since $g_0=0$ corresponds to the pure case
it is intuitively clear that the parameter $g_0$ is an increasing
function of the degree of disorder. For weak disorder this implies
that $g_0$ is very small and therefore, due to the exponential dependence,
that the asymptotic region in $t$ is extremely narrow.

Strictly speaking the model function (\ref{eq:chi_model}) 
should resemble the ``true'' susceptibility only
for weak disorder, but it is commonly believed that it is a 
reasonable qualitative approximation also for strong disorder. 
To relate the parameter
$g_0$ at least heuristically to the ratio $J_2/J_1$ we used the weak
disorder result $g_0 = c_2 a^2/(1+a b)^2$, where $c_2 = 1 - x$
(with $x$ as defined in eq.~(\ref{eq:P})) is assumed to be small, $c_2 \ll 1$,
i.e., the analytic calculation assumes that there are only few $J_2$-bonds
in a background of $J_1$-bonds. The parameters $a$ and $b$ are given by
$a = (v'_c - v^{(0)}_c)/v^{(0)}_c$ and
$b = v^{(0)}_c/2\sqrt{2}$, where $v^{(0)}_c = \tanh(\beta^{(0)}_c J_1)
=\sqrt{2}-1$ and $v'_c = \tanh(\beta^{(0)}_c J_2)$, with 
$\beta^{(0)}_c$ denoting the inverse critical temperature of the 
{\em pure\/} system with all $J_{ij} = J_1$. Of course, employing this
formula to the present case with $c_2 = 1/2 = x$ is a bold step which even
creates an ambiguity since the exact symmetry $J_1 \leftrightarrow J_2$ for
$x=1/2$ is violated. For weak disorder ($J_2/J_1 \approx 1$), however, the 
inconsistency turns out to be very mild. For $J_2/J_1 = 1.2$ we obtain
$g_0 = 0.013700\dots$, and for $J_2/J_1 = 1/1.2$ we find a slightly
smaller value of $g'_0 = 0.011958\dots$. This shows that for weak disorder 
($J_2/J_1 = 1.2$, $g_0 \approx 0.013$) the asymptotic region is bounded
by $t \ll \exp(-\pi/4 g_0) \approx \exp(-1/0.017) \approx 10^{-26}$,
and thus explains why it is so difficult to observe the asymptotic critical
behavior in the weak-disorder limit. For $J_2/J_1 = 1.5$ and $1/1.5$ the 
corresponding numbers are $g_0 = 0.070700\dots$ and $g'_0 = 0.052889\dots$, 
leading to a bound of the order of $t \ll 10^{-5}$.

\begin{figure}[bhp]
\vskip 7.0truecm
\includegraphics{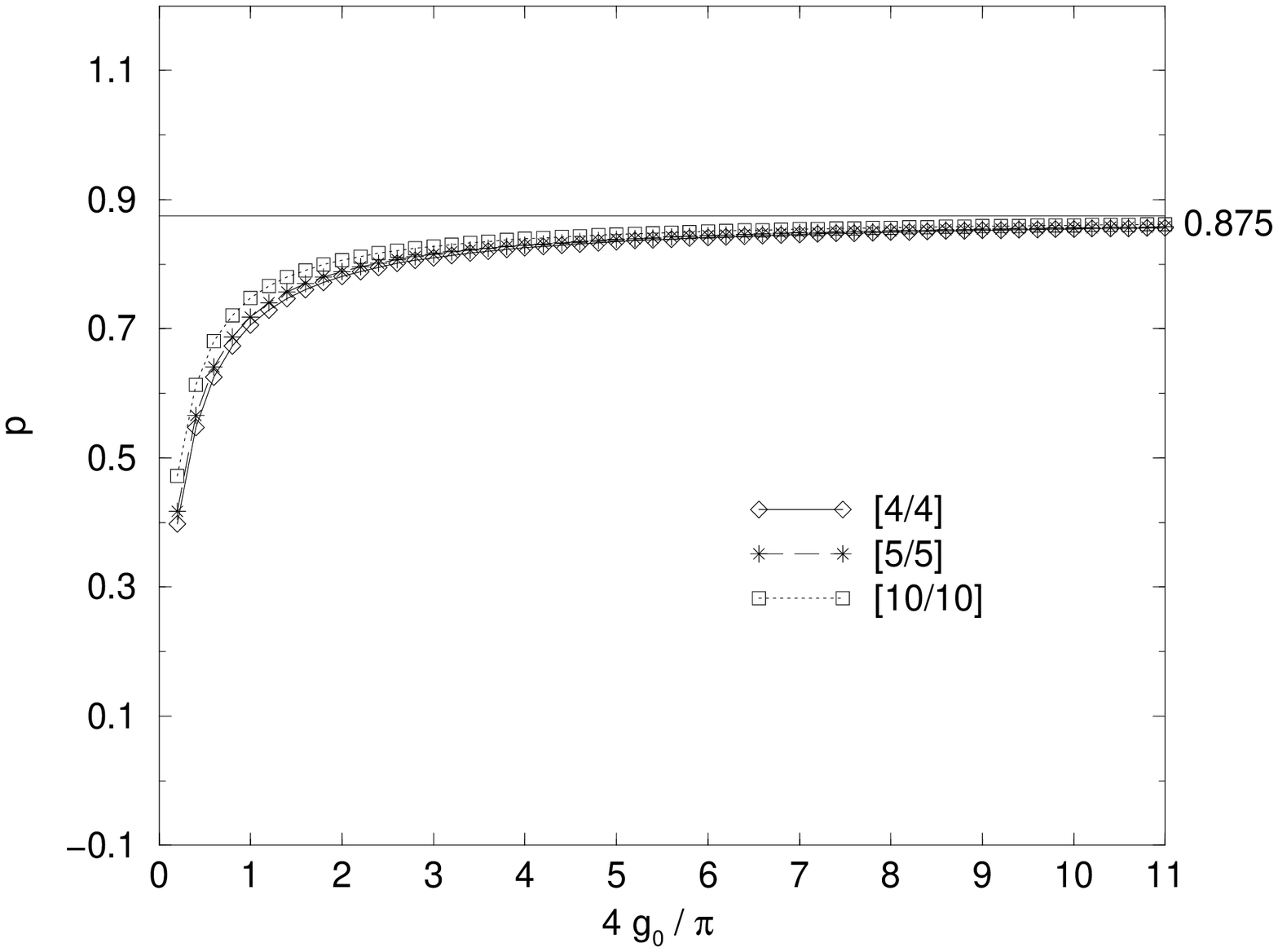}
\caption[a]{{\em Analysis of the series expansion of the model
function (\ref{eq:chi_model}), using the ansatz 
$\chi_{\rm model} \propto t^{-7/4} |\ln t|^p$. The legend indicates 
the three different Pad\'e approximants shown. The horizontal line
shows the exact value $p = 7/8 = 0.875$.
}}
\label{fig:chi_model}
%\end{figure}
%
%\clearpage\newpage
%
%-------------------------------------------------------------------
%                         Figure 4
%-------------------------------------------------------------------
%
%\begin{figure}[bhp]
\vskip 8.0truecm
\includegraphics{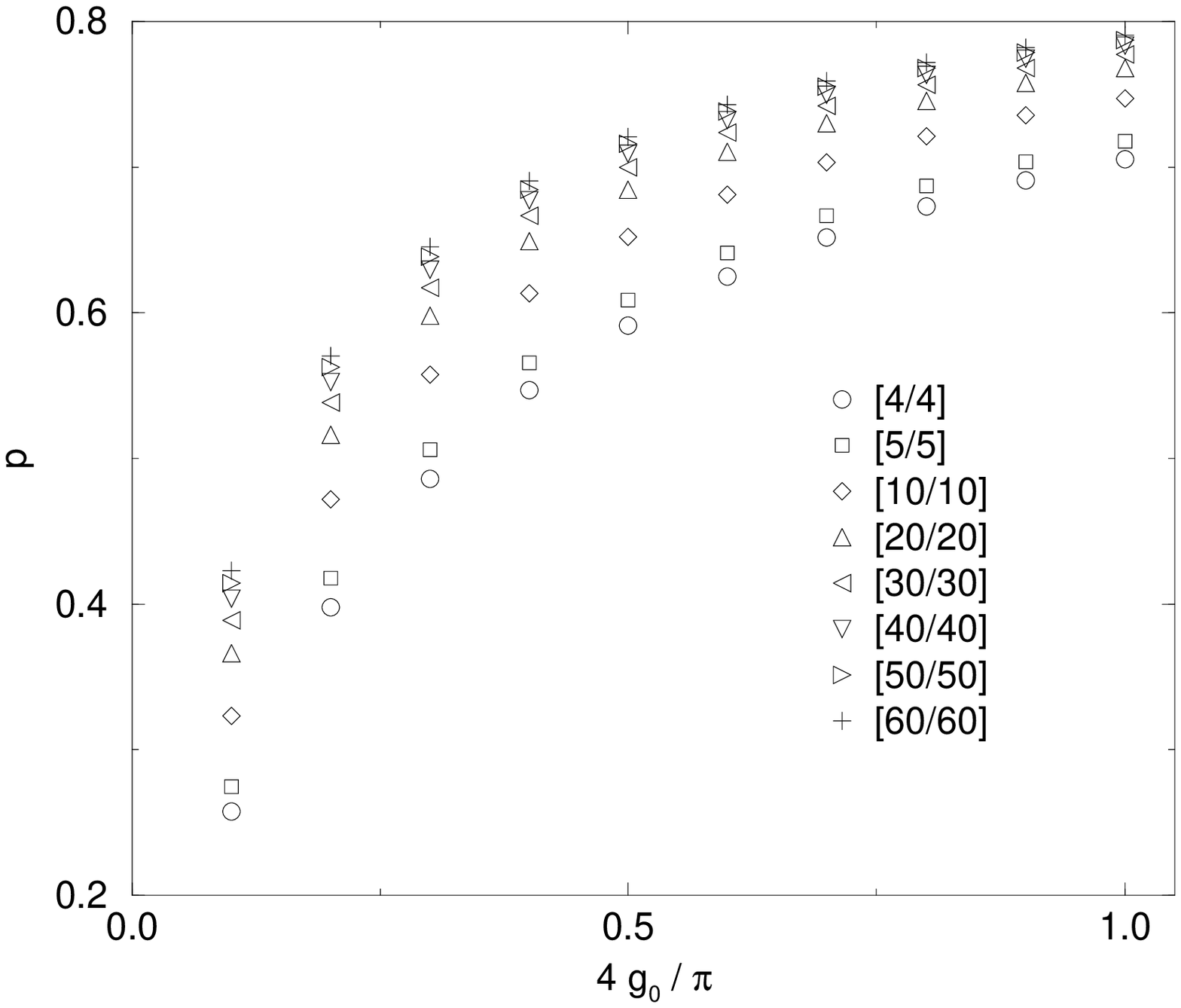}
\caption[a]{{\em Convergence behavior of the model series
(\ref{eq:chi_model})
for the susceptibility with increasing order at fixed parameter $g_0$. The
symbols $[L/M]$ denote the various Pad\'e approximants used.
}}
\label{fig:chi_model_4-60}
\end{figure}

Using a symbolic computer program it is straightforward to generate the
high-temperature series expansion of the model function (\ref{eq:chi_model})
to any desired order. Applying the same analysis techniques as used for the
``true'' susceptibility series we obtained the results shown in 
Fig.~\ref{fig:chi_model}. If we truncate the model series at low order
we observe qualitatively the same crossover effect as for the ``true'' series.
Here we are sure, however, that this must be a pure artifact of the truncation
of the model series at a finite order. We also see that the approach of
the asymptotic limit of $p=7/8$ as a function of the degree of disorder
is faster if we consider a longer series (21 terms). 
It is,
however, somewhat discouraging (even though understandable in view of the 
exponential dependence of the critical regime on $g_0$) 
that at a fixed $g_0$ the convergence of the series with increasing order
is quite
slow. For example, at $4 g_0/\pi = 1$ we obtained $p=0.7056$ with the 
Pad\'e approximant $[4/4]$, 0.7178 ($[5/5]$), 0.7474 ($[10/10]$), 
0.7682 ($[20/20]$), 0.7777 ($[30/30]$), 0.7834 ($[40/40]$),
0.7875 ($[50/50]$), and 0.7905 ($[60/60]$).
The convergence behavior for this example and other small values of the 
parameter $g_0$ can be visually inspected in Fig.~\ref{fig:chi_model_4-60}.

\paragraph{Specific heat:}
Series analyses for the specific heat are usually more difficult than for
the susceptibility. This is especially pronounced for the Ising model
on loose-packed lattices where all odd powers of $\beta$ vanish because
of symmetry. Consequently our specific-heat series consists only of four
non-trivial terms (see Table~\ref{tab:free.series}). We nevertheless tried
an analysis with the ansatz 
\begin{equation}
C \propto |\ln t|^q, 
\end{equation}
using the method
``Ln-Pad\'e''. The exponent $q$ is an effective exponent whose value may, or
 may not be constant.

\begin{figure}[bhp]
\vskip 7.0truecm
\includegraphics{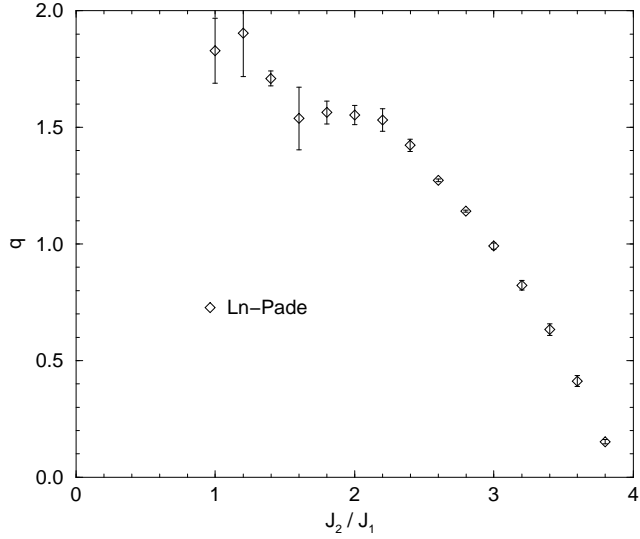}
\caption[a]{{\em Analysis of the specific-heat series using
the Ln-Pad\'e method.
}}
\label{fig:c_lnPade}
\end{figure}

The resulting dependence of the exponent $q$ on the 
ratio $J_2/J_1$ is shown in Fig.~\ref{fig:c_lnPade}. While the
quantitative agreement with the exactly known pure case is certainly
not convincing, we do see at least a qualitative trend to smaller values
of $q$ with increasing strength of the disorder (increasing ratios $J_2/J_1$),
i.e., the singularity of the specific heat becomes apparently 
weaker for stronger disorder. This may be taken as an indication that the 
true singularity is of the log-log type (\ref{eq:C_DD}) 
as predicted by both, DD and SSL. A recent numerical study \cite{reis2} for 
$J_2/J_1 = 4$
using transfer-matrix methods also observed a behavior in between log and 
log-log type. These findings are in contradiction to the 
claim \cite{kim1} for a slightly different disordered system (quenched, 
random site-dilution) that the specific heat stays finite at $T_c$, as 
theoretically suggested in Ref.~\cite{ziegler} (see also 
Ref.~\cite{comments}).

\begin{figure}[bhp]
\vskip 7.0truecm
\includegraphics{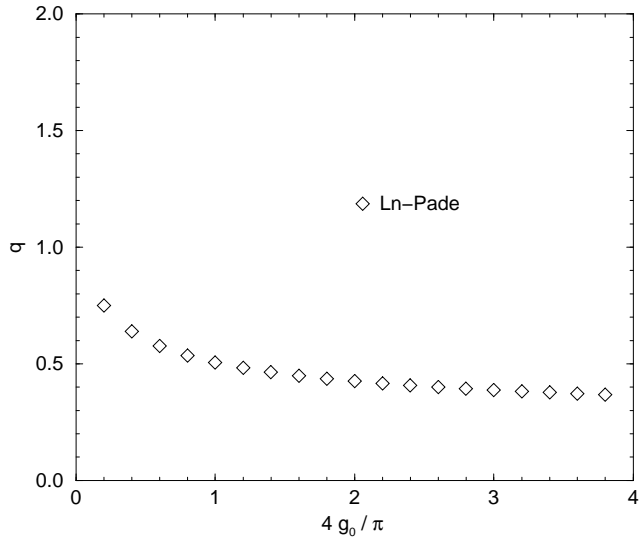}
\caption[a]{{\em Analysis of the series expansion of the model 
specific-heat using the Ln-Pad\'e method.
}}
\label{fig:c_model}
\end{figure}

Again
we have tried to justify our interpretation by considering a model function,
\begin{equation}
C_{\rm model} = \frac{1}{g_0} \ln\left[1 + \frac{4 g_0}{\pi} \ln(1/\dot{t})\right]. 
\end{equation}
By applying precisely the same type of analysis to the series expansion 
of the model specific-heat we obtained the results displayed in
Fig.~\ref{fig:c_model}, which show qualitatively the same trend 
of decreasing $q$ as a function of $J_2/J_1$ as the data 
in Fig.~\ref{fig:c_lnPade}. 

%
%---------------------------------------------------------
            \section{Discussion} \label{sec:conclusions}
%---------------------------------------------------------
%
The main results of our high-temperature series analysis are shown in
Fig.~\ref{fig:chi_SSL} which provide at least for strong
disorder (large $J_2/J_1$) compelling evidence that the singularity of the
susceptibility is properly described by $\chi \propto t^{-7/4} |\ln t|^p$,
with $p= 7/8 = 0.875$, as theoretically predicted by 
SSL [3-6].
% \cite{shalaev1,shalaev2,shankar,ludwig}. 
The analysis of the
model susceptibility (\ref{eq:chi_model}) in Figs.~\ref{fig:chi_model}
and \ref{fig:chi_model_4-60}
clearly shows that the apparent variation of $p$ with the strength of
disorder is an artifact caused by the truncation of the series expansions
at a finite order. We, therefore, emphasize that the apparent crossover
from weak to strong disorder does {\em not} imply that the universality
class of the random-bond Ising model changes continuously with the
strength of disorder. 

Let us finally make a few comments on previous Monte Carlo simulations
of this model on large but finite square lattices. 
With the finite-size scaling analysis of  
% Refs.~\cite{andreichenko1,wang1,wang2,selke0,selke1,reis1},
Refs.~[8-12, 16]
it is conceptually impossible to detect
the multiplicative logarithmic correction of the SSL prediction
(\ref{eq:chi_SSL}). The reason is that the SSL theory also predicts
a logarithmic correction for the scaling behavior
of the correlation length, $\xi \propto t^{-1} |\ln t|^{1/2}$. 
In the finite-size scaling behavior the two logarithms thus cancel and one
ends up with a pure power-law, $\chi \propto L^{\gamma/\nu} = L^{7/4}$, where
$L$ is the linear lattice size. Thus {\em only\/} the SSL prediction for 
$\gamma/\nu$ can be tested in finite-size scaling analyses. 
Wang {\em et al.\/} \cite{wang1,wang2}
obtained for $J_2/J_1 = 4$ and 10 an estimate of $\gamma/\nu = 1.7507
\pm 0.0014$, and also the results of Reis {\em et al.\/} \cite{reis1} 
at $J_2/J_1 = 2$, 4, and 10 are consistent with $\gamma/\nu = 1.75$. Among 
the two alternatives, the theories of DD and SSL, these estimates thus 
provide evidence in favor of SSL. Notice, however, that a numerical estimate 
of $\gamma/\nu \approx 1.75$ 
would also be expected for the {\em pure} two-dimensional Ising model. 
For the specific heat the situation is conceptually clearer. Here the 
theoretically expected scaling behavior (\ref{eq:C_DD}), as predicted by both,
DD and SSL, translates into a double-logarithmic finite-size scaling behavior,
$C = C_0 + C_1 \ln(1 + b \ln L)$, which is different from that of the pure
case where $C = C_0 + \ln L$. In the numerical work of 
Wang {\em et al.\/} \cite{wang1,wang2}, employing lattice sizes up 
to $L=600$, this difference in the 
asymptotic behavior is clearly observed for $J_2/J_1 = 10$, while for
$J_2/J_1 = 4$ the behavior is in between log and log-log type, similar to the
findings reported in a recent transfer-matrix study \cite{reis2} for the
same coupling-constant ratio.
For the specific heat these latest finite-size scaling analyses are thus about
as conclusive as our series analyses in Fig.~\ref{fig:c_lnPade}.

Another set of numerical data that can discriminate between the 
predictions of DD and SSL
comes from direct simulations of the temperature dependence of the
magnetization $m$ and of the susceptibility $\chi$ for $J_2/J_1 = 4$
\cite{wang2,talapov1}. Assuming in the analysis a pure power law with an
{\em effective\/} exponent (i.e., ignoring the logarithmic correction), one 
observes an overshooting of the effective exponent to values larger than the
prediction by SSL. 
As discussed above (recall eq.~(\ref{eq:chi_eff})) this may be taken as an 
indication of a multiplicative logarithmic correction term.
For example, Talapov and Shchur \cite{talapov1} 
obtained for $J_2/J_1 = 4$ from least-squares
fits to $\chi \propto t^{-\gamma_{\rm eff}}$ an effective
exponent of $\gamma_{\rm eff} \approx 7/4 + 0.135 = 1.885$. This value is 
quite close to our series estimate of $\gamma_{\rm eff} = 2.019 \pm 0.024$ 
for $J_2/J_1 = 4$,
if the pure power-law ansatz is used (cp.\ Fig.~\ref{fig:chi_DD}). Wang {\em
et al.\/} \cite{wang2} furthermore confirmed that their data is compatible 
with the SSL ansatz, $\chi(t) = \chi_0 t^{-7/4} (1 + a t) [ 1 + 
b \ln(1/t)]^{7/8}$, supplemented by a correction to scaling term 
$(1 + a t)$ (and similarly for $m$; for a recent confirmation, 
see Ref.~\cite{ssli}). In these fits both exponents are kept fixed at
their predicted values, and $\chi_0$, $a$, and $b$ are free parameters. 
In contrast to our series analysis, however, no quantitative estimates of the 
exponent of the logarithmic correction have been reported in 
Ref.~\cite{wang2}.
While the simulation results certainly indicate that among the two
conflicting theories of DD and SSL, 
the SSL prediction is more likely to be correct, 
it is still fair to conclude that also this set of simulations has not yet
unambiguously identified the multiplicative logarithmic correction term.

Monte Carlo simulations of systems with quenched, random disorder require
an enormous amount of computing time because many realisations have to be
simulated for the quenched average. For this reason it is hardly possible
to scan a whole parameter range. 
Using high-temperature series expansions, on the other hand, one can obtain
closed expressions in several parameters (such as the dimension $d$, $x$,
$J_2/J_1$, \dots) up to a certain order in the
inverse temperature $\beta = 1/k_BT$.
Here the infinite-volume limit is always implied and the quenched, random
disorder can
be treated exactly. By analysing the resulting series, the critical behavior
of the random-bond system can hence in principle be monitored as a continuous
function of several parameters. This is a big advantage over Monte Carlo
simulations which 
usually can only yield a
rather small parameter range in one set of simulations. The
caveat of the series-expansion approach is that the available series 
expansions for the random-bond
Ising model are still relatively short
(at any rate much shorter than for pure systems). This introduces
systematic errors of the resulting estimates for critical exponents which
are difficult to control. The obvious way out is trying to extend the series
expansions as far as possible. This, however, would be extremely cumbersome 
since the number
of algebraic manipulations necessary to calculate the series coefficients
blows up dramatically with the order of the series (usually at least
exponentially) and, therefore, has to be left for future work. 
%
%---------------------------------------------------------
            \section*{Acknowledgements}
%---------------------------------------------------------
%
We wish to thank Kurt Binder for many helpful discussions
and his constant interest in this project. JA and WJ are grateful to
Walter Selke for a critical discussion of previous numerical
simulation results, and thank Dietrich Stauffer for comments on 
the manuscript.
WJ acknowledges support from the Deutsche Forschungsgemeinschaft through 
a Heisenberg fellowship.
Partial support of JA and WJ from the German-Israel-Foundation (GIF) under 
contract No. I-0438-145.07/95 is also gratefully acknowledged.
%
%-------------------------------------------------------------------
%

\clearpage\newpage
%
%-------------------------------------------------------------------
%                             Table 1
%
\newlength{\digitwidth} \settowidth{\digitwidth}{\rm 0}
\catcode`?=\active \def?{\kern\digitwidth}
%-------------------------------------------------------------------
%
\begin{table}
\begin{center}
\caption[a]{\label{tab:free.series}
Coefficients $a_n$ of the high-temperature series expansion of the 
free energy per site, 
$-\beta f = \ln[2\cosh(\beta J_1)\cosh(\beta J_2)] + \sum_n a_n k^n$, 
with $k = 2 \beta J_1$.\\
}
\begin{tabular}{|r|r|r|r|r|}
\hline
 \multicolumn{1}{|c}{$n$}          &
 \multicolumn{1}{|c}{$J_2/J_1=1$}  &
 \multicolumn{1}{|c}{$J_2/J_1=2$}  &
 \multicolumn{1}{|c}{$J_2/J_1=4$}  &
 \multicolumn{1}{|c|}{$J_2/J_1=10$} \\
\hline & & & & \\[-0.3cm]
 4 & $\frac{1}{16}$ & $\frac{81}{256}$ & $\frac{625}{256}$ & 
     $\frac{14641}{256}$   \\[0.2cm]
 6 & $\frac{1}{96}$ & $\frac{81}{2048}$ & $-\frac{18125}{6144}$ & 
     $-\frac{5343965}{6144}$   \\[0.2cm]
 8 & $\frac{17}{2560}$ & $\frac{33671}{327680}$ & $-\frac{26293}{65536}$  & 
     $-\frac{990135929}{327680}$   \\[0.2cm]
10 & $\frac{1907}{483840}$ & $\frac{3437297}{27525120}$ & 
     $\frac{1057390637}{49545216}$  & 
     $\frac{16514750542133}{49545216}$   \\[0.2cm]
\hline
\end{tabular}
\end{center}
\end{table}
%
%-------------------------------------------------------------------
%                             Table 2
%-------------------------------------------------------------------
%
\begin{table}
\begin{center}
\caption[a]{\label{tab:chi.series}
Coefficients $b_n$ of the high-temperature series expansion of the 
susceptibility per site, $\chi = 1 + \sum_n b_n k^n$, 
with $k = 2 \beta J_1$.\\
}
\begin{tabular}{|r|r|r|r|r|}
\hline
 \multicolumn{1}{|c}{$n$}          &
 \multicolumn{1}{|c}{$J_2/J_1=1$}  &
 \multicolumn{1}{|c}{$J_2/J_1=2$}  &
 \multicolumn{1}{|c}{$J_2/J_1=4$}  &
 \multicolumn{1}{|c|}{$J_2/J_1=10$} \\
\hline & & & & \\[-0.3cm]
 1 & 2 & 3 & 5 & 11   \\[0.2cm]
 2 & 3 & $\frac{27}{4}$ & $\frac{75}{4}$ & $\frac{363}{4}$ \\[0.2cm]
 3 & $\frac{13}{3}$ & $\frac{231}{16}$ & $\frac{3115}{48}$  & 
     $\frac{31933}{48}$   \\[0.2cm]
 4 & $\frac{23}{4}$ & $\frac{1809}{64}$ & $\frac{13025}{64}$  & 
     $\frac{277937}{64}$   \\[0.2cm]
 5 & $\frac{451}{60}$ & $\frac{69337}{1280}$ & $\frac{471185}{768}$  & 
     $\frac{101248147}{3840}$   \\[0.2cm]
 6 & $\frac{191}{20}$ & $\frac{515871}{5120}$ & $\frac{1823875}{1024}$  & 
     $\frac{768499919}{5120}$   \\[0.2cm]
 7 & $\frac{30283}{2520}$ & $\frac{79576207}{430080}$ &
     $\frac{1302083479}{258048}$  & $\frac{1034056024661}{1290240}$  \\[0.2cm]
 8 & $\frac{100003}{6720}$ & $\frac{191638233}{573440}$ &
     $\frac{4823704415}{344064}$  & $\frac{7079050432267}{1720320}$  \\[0.2cm]
 9 & $\frac{3318601}{181440}$ & $\frac{587805509}{983040}$ &
     $\frac{203928262469}{5308416}$  & 
     $\frac{3850544162365417}{185794560}$ \\[0.2cm]
10 & $\frac{3369629}{151200}$ & $\frac{48645511629}{45875200}$ &
     $\frac{5160699783175}{49545216}$  & 
     $\frac{126985060534491247}{1238630400}$ \\[0.2cm]
11 & $\frac{269543489}{9979200}$ & $\frac{101837138460677}{54499737600}$ &
     $\frac{9157142004160957}{32699842560}$  &
     $\frac{1069481408075459203}{2123366400}$   \\[0.2cm]
\hline
\end{tabular}
\end{center}
\end{table}
\clearpage\newpage
%
%-------------------------------------------------------------------
%                         Figure 1
%-------------------------------------------------------------------
%
%
%\clearpage\newpage
%
%-------------------------------------------------------------------
%                         Figure 5 and 6
%-------------------------------------------------------------------
%
\end{document}